\documentclass{moriond}

\usepackage{float}

\def\be{\begin{equation}}
\def\ee{\end{equation}}
\def\bea{\begin{eqnarray}}
\def\eea{\end{eqnarray}}

\newcommand{\Photo}{}

\begin{document}
\vspace*{4cm}
\title{Observation of vector charmoniumlike states and search for $Z_{cs}$ in $e^+ e^- \to K^+ K^- J/\psi$ at BESIII}

\author{Hang Zhou (on behalf of the BESIII Collaboration)}

\address{Research Center of Particle Physics and Technology, Shandong University, \\
72 Binhai Road, Qingdao 266237, P.~R.~China}

\maketitle\abstracts{
We present the recent measurements of the $e^+e^-\to K^+K^- J/\psi$ process carried out by the BESIII experiment. A new decay mode $Y(4230)\to K^+K^- J/\psi$ is firstly identified by BESIII. Furthermore, two vector charmoniumlike states, $Y(4500)$ and $Y(4710)$, are reported for the first time in the energy-dependent line shape of  $e^+e^-\to K^+K^- J/\psi$ cross section.
In additon, an investigation of the $K^\pm J/\psi$ system is conducted to explore charged charmoniumlike states.
}

\section{Introduction}
Since their discovery, charmonia have been instrumental in advancing our understanding of strong interactions and testing QCD. Our knowledge of charmonium system below the open charm threshold is well-developed. The charmonium spectrum is well described by a potential model~\cite{Kwong:1987mj}, with excellent agreement between theories and experiments. Above the open-charm threshold, however, a comprehensive understanding is still lacking. Since 2003, several unexpected states, such as $X(3872)$, $Y(4260)$, and $Z_c(3900)$, have been observed. These particles do not match the predictions of potential models, leading to the proposal of configurations beyond the conventional quark model, such as multi-quark states, hadronic molecules, or hybrids~\cite{Brambilla:2019esw,Chen:2022asf}. 
The persistent mystery surrounding these states has triggered tremendous efforts to crack their nature and hunt new ones over the past two decades, and dedicated studies are still ongoing.

The Beijing Spectrometer (BESIII) experiment~\cite{BESIII:2022wjl} is a symmetric $e^+e^-$ collisions experiment running at the Beijing Electron Positron Collider II (BEPCII) with a center-of-mass (c.m.) energy ranging from 2.0 to 4.95 GeV. The BESIII has accumulated the world's largest $e^+e^-$ collisions data sample in this energy range, which provides an ideal platform to explore the exotic particles.



\section{Observation of vector charmoniumlike states in $e^+e^-\to K^+ K^- J/\psi$}

In recent years, numerous vector states, e.g.~$Y(4230)$, $Y(4320)$, $Y(4360)$, and $Y(4660)$, have been found above open-charm threshold, where the well established $\psi$ states, $\psi(3770)$, $\psi(4040)$, $\psi(4160)$, and $\psi(4415)$, are also present~\cite{Workman:2022ynf}. The plethora of vector states in this energy region appears to pose a challenge to potential models.
In addition, a striking feature of these newly discovered $Y$ particles, in contrast to the vector charmonia, is their absence in the $R=\sigma(e^+e^-\to hadrons)/\sigma(e^+e^-\to \mu^+\mu^-)$ measurement. As their identities
are still unknown, further experimental measurements are necessary.

\begin{figure}[htpb]
    \centering
    \includegraphics[width=0.6\linewidth]{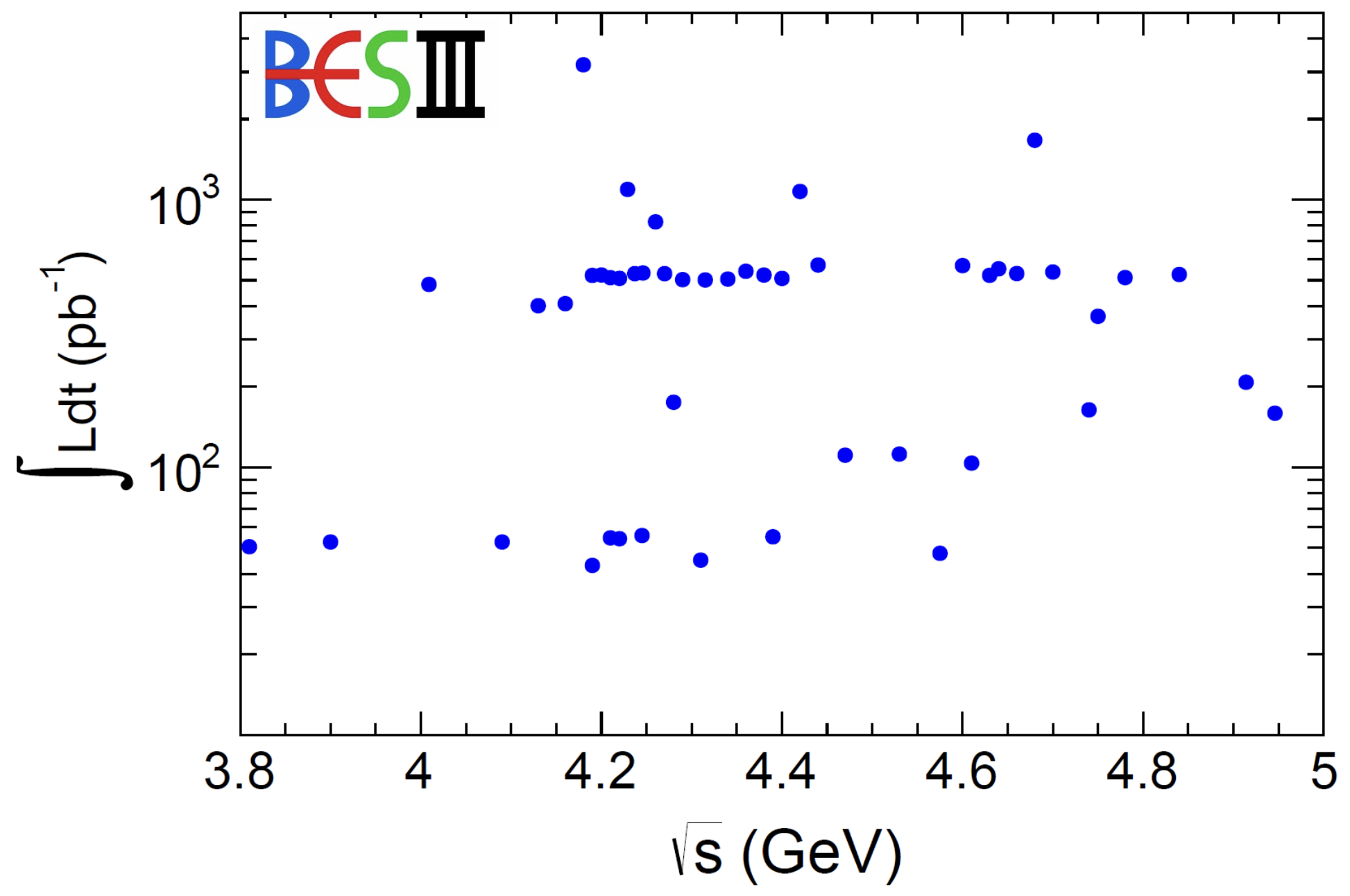}
    \caption{The integrated luminosity of the data sets collected by BESIII at c.m.~energies between 3.8 and 5.0 GeV.}
    \label{fig:data_XYZ}
\end{figure}

The vector states with $J^{PC}=1^{--}$ are directly accessible via the $e^+e^-$ collisions. Since 2013, BESIII has been conducting intensive high luminosity energy scans between 4 and 5 GeV, which allows precise studies of vector states in this energy region. Figure~\ref{fig:data_XYZ} shows the data sets accumulated by BESIII at c.m.~energies ranging from 3.8 to 5.0 GeV, corresponding to a total integrated luminosity of 22~$\rm{fb}^{-1}$.

\begin{figure}[htbp]
    \centering
    \includegraphics[height=5.5cm]{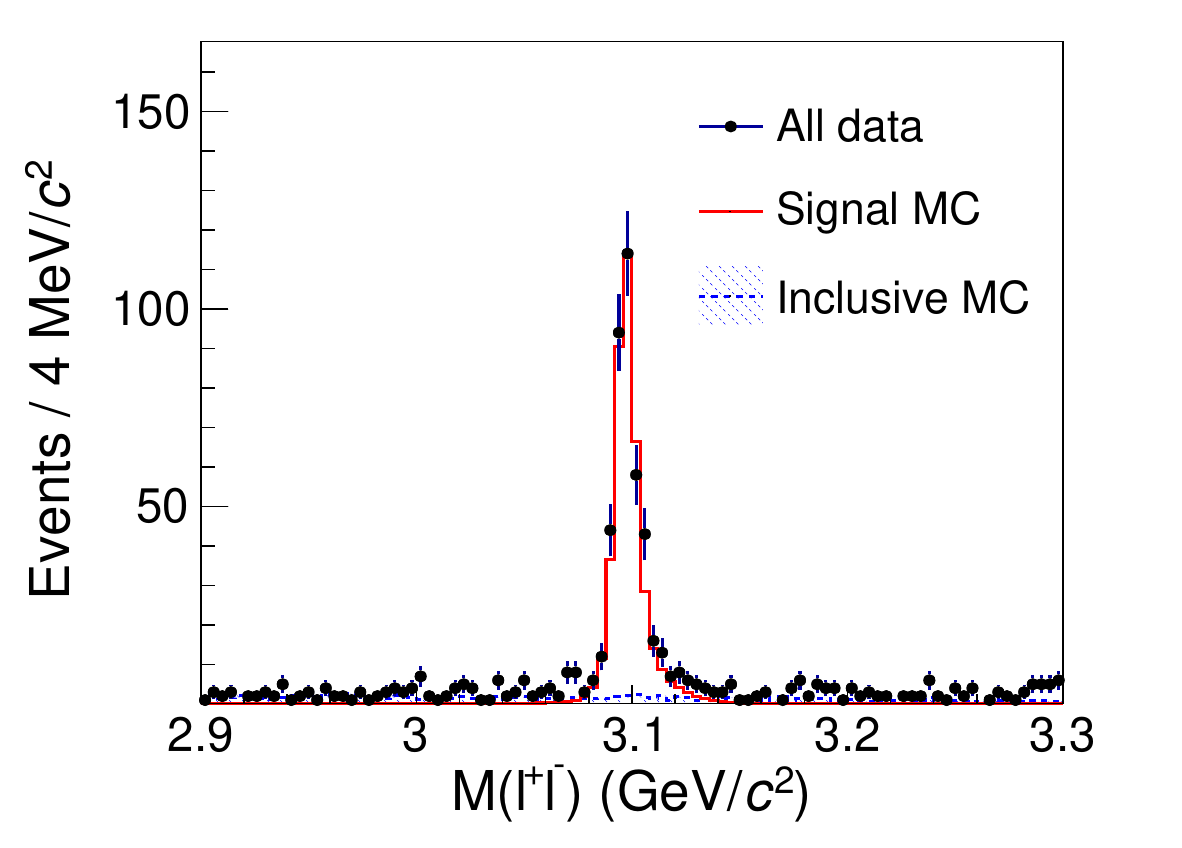}
    \includegraphics[height=5.5cm]{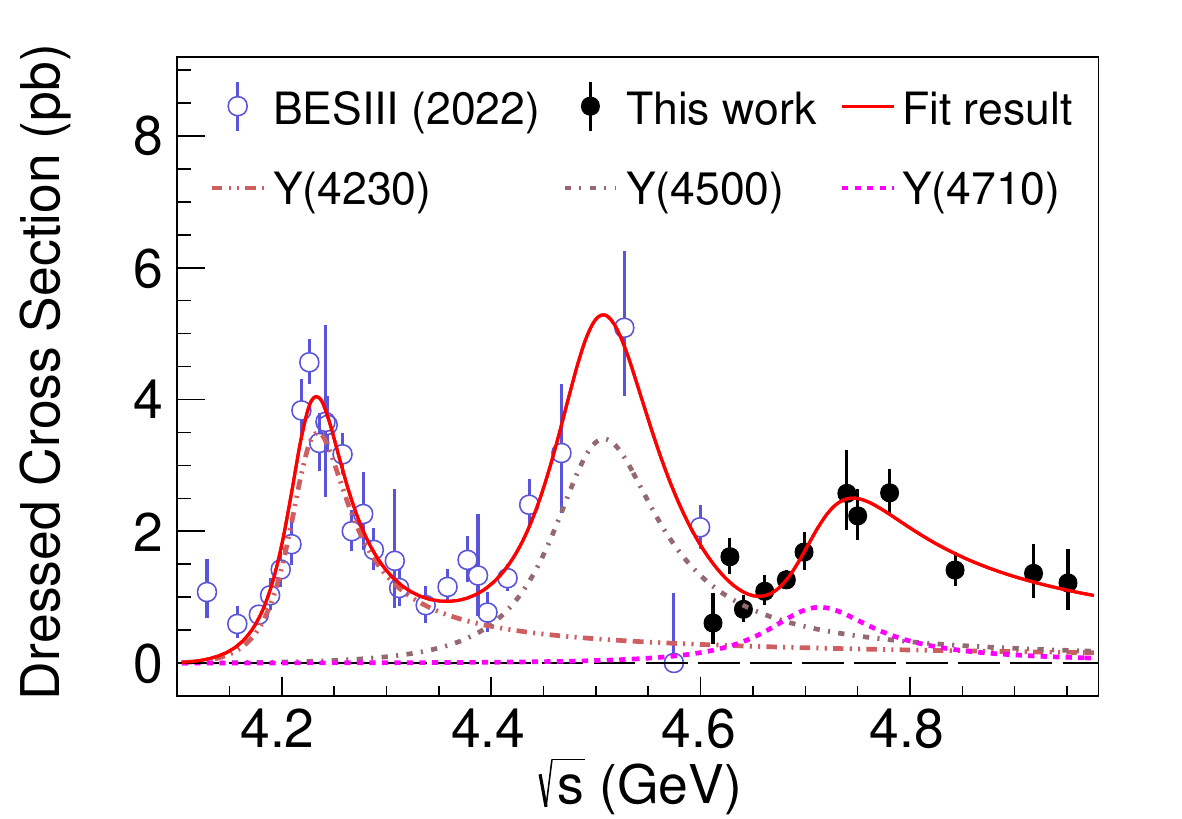}
    \caption{
    Left: The invariant mass distribution of $\ell^+\ell^-$. Right: The production cross section of $e^+e^-\to K^+K^- J/\psi$ at BESIII~\protect\cite{BESIII:2023wqy}.
    }
    \label{fig:mll-xs}
\end{figure}

With these data sets, the $e^+e^- \to K^+ K^- J/\psi$ cross sections are precisely measured from its threshold to 5 GeV~\cite{BESIII:2022joj,BESIII:2023wqy}. In order to achieve a much lower background level and improve the statistic, both full reconstruction ($K^+K^-\ell^+\ell^-$) and partial reconstruction ($K^\pm K^\mp_\mathrm{miss}\ell^+\ell^-$) methods are applied for the signal reconstruction. This approach is made possible owing to the clear kinematics of the process and the effective background control at BESIII.
Figure~\ref{fig:mll-xs} (left) shows the invariant mass distribution of lepton pair, $M(\ell^+\ell^-)$, the $J/\psi$ peak is clearly reconstructed with very low background level. Figure~\ref{fig:mll-xs} (right) shows the cross sections of $e^+e^- \to K^+ K^- J/\psi$ measured by BESIII, three significant structures are observed. The first one, with mass and width measured to be ($4225.3 \pm 2.3 \pm 21.5$)~$\mathrm{MeV}/c^2$ and ($72.9 \pm 6.1 \pm 30.8$)~MeV, respectively, is consistent with the $Y(4230)$~\cite{BESIII:2016bnd}. Additionally, two structures in the cross sections at around 4.5 GeV and 4.75 GeV, denoted as $Y(4500)$ and $Y(4710)$, are observed for the first time with statistical significance exceeding $5\sigma$. The mass and width of $Y(4500)$ are measured to be ($4484.7 \pm 13.3 \pm 24.1$)~$\mathrm{MeV}/c^2$ and ($111.1 \pm 30.1 \pm 15.2$)~MeV,
respectively. The $Y(4710)$ has a mass and width of ($4708^{+17}_{-15} \pm 21$)~$\mathrm{MeV}/c^2$ and ($126^{+27}_{-23} \pm 30$)~MeV,
respectively. The $Y(4710)$ is also observed in $e^+e^- \to K_S^0 K_S^0 J/\psi$ process with a statistical significance of $4.2\sigma$~\cite{BESIII:2022kcv}.

At present, a common understanding of these vector charmoniumlike states has not been reached. Someone suggest that they could be exotic candidates like $c\bar{c}g$ hybrid, tertraquark, or molecule, while others consider they are simply charmonium states. For instance, in reference~\cite{Wang:2022jxj}, a charmonium mixing scheme is proposed to address the abundance of vector states observed. 

\section{Search for $Z_{cs}$ state in $K^\pm J/\psi$ system}

The charged charmoniumlike state, comprising at least four quarks $c\bar{c}q\bar{q}'$, stands out as one of the most promising exotic particles. The first charged charmoniumlike state, $Z_c(3900)$ with minimal quark content of $c\bar{c}u\bar{d}$, was discovered in $\pi J/\psi$ by BESIII~\cite{BESIII:2013ris}. At that time, the existence of its SU(3) partner, $Z_{cs}$ with $c\bar{c}u\bar{s}$ content, was anticipated. 
Until 2021, BESIII first announced the observation of a near-threshold structure, denoted as $Z_{cs}(3985)$, in the $K^+$ recoil-mass spectra from the $e^+ e^- \to K^+(D_s^- D^{*0} + D^{*-}_s D^0)$ process~\cite{BESIII:2020qkh}. This state has a mass and width of 3982.5~$\mathrm{MeV}/c^2$ and 13.8 MeV, respectively. 
In the same year, LHCb reported two structures, named $Z_{cs}(4000)$ and $Z_{cs}(4220)$, in the $K^\pm J/\psi$ system in the $B^\pm\to\phi K^\pm J/\psi$ decay~\cite{LHCb:2021uow}. The $Z_{cs}(4000)$ has a mass of 4003~$\mathrm{MeV}/c^2$ and a width of 131 MeV,
and the $Z_{cs}(4220)$ has a mass of 4216 $\mathrm{MeV}/c^2$ and a width of 233 MeV. Interestingly, the masses of $Z_{cs}(3985)$ and $Z_{cs}(4000)$ are quite similar, while their widths are notably different. 
There are ongoing debates on whether they are the same state or not, and maybe one of them is the strange partner of $Z_c(3900)$.   
Very recently, both BESIII~\cite{BESIII:2022qzr} and LHCb~\cite{LHCb:2023hxg} reported evidence of neutrual partners of these $Z_{cs}$ states, which indicates they form isospin doublets.

\begin{figure}[htbp]
    \centering
    \includegraphics[height=6cm]{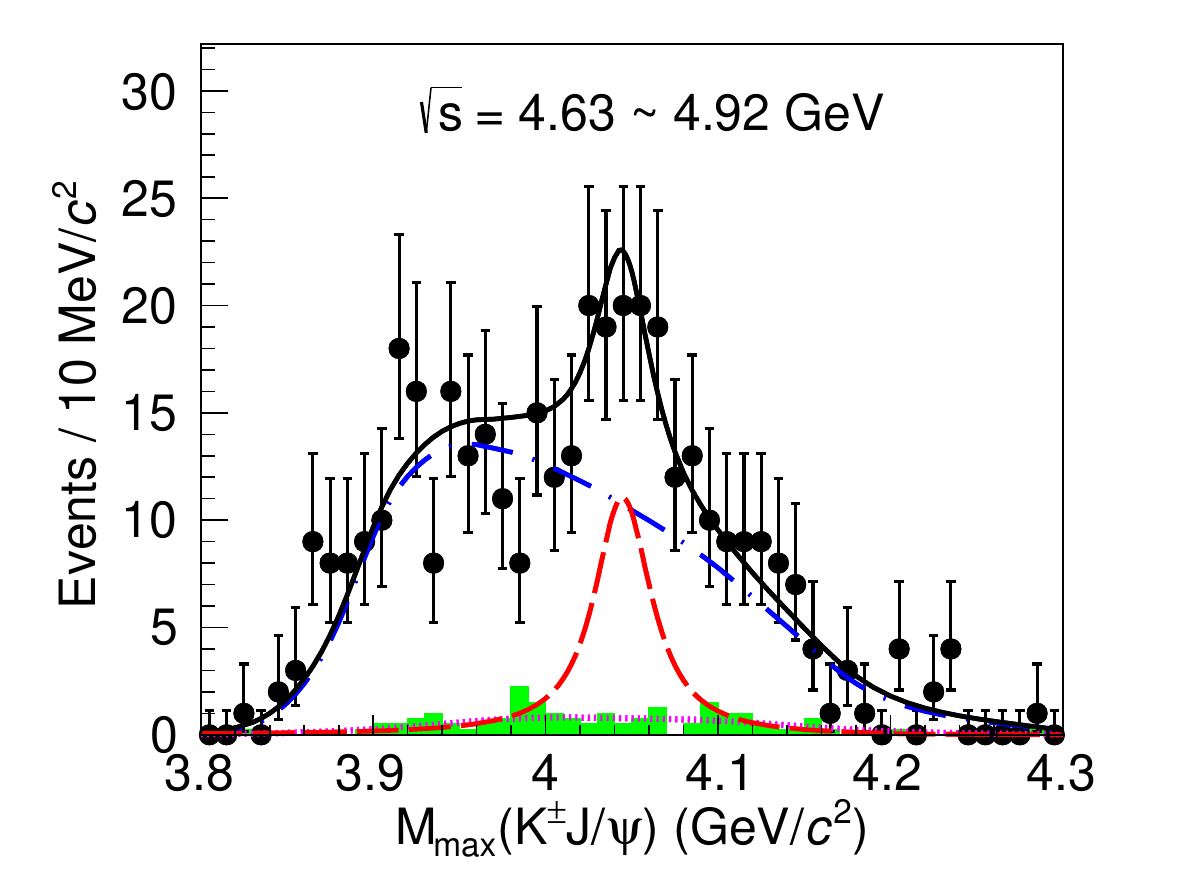}
    \caption{The distribution of the maximum of invariant masses of $K^+ J/\psi$ and $K^- J/\psi$ combinations~\protect\cite{BESIII:2023wqy}.}
    \label{fig:mkjpsi}
\end{figure}

Following these observations by BESIII and LHCb, searching for $Z_{cs}$ states in $e^+e^- \to K^+ K^- J/\psi$ is highly sought after. Using a data sample with an integrated luminosity of 5.6~$\mathrm{fb}^{-1}$, the $K^\pm J/\psi$ system is investigated in the $e^+e^- \to K^+ K^- J/\psi$ process~\cite{BESIII:2023wqy}. Figure~\ref{fig:mkjpsi} displays the distribution of the maximum of invariant masses of $K^+ J/\psi$ and $K^- J/\psi$ combinations, $M_\mathrm{max}(K^\pm J/\psi)$. There is no evidence of $Z_{cs}(3985)$ or $Z_{cs}(4000)$, but a minor peak with a significance of $2.3\sigma$ is seen between $D^*_sD/D^*D_s$ and $D_s^*D^*$ mass thresholds. The ratio of branching fractions $\frac{\mathcal{B}[Z_{cs}(3985)^{+}\rightarrow K^+ J/\psi]}{\mathcal{B}[Z_{cs}(3985)^{+}\rightarrow (\bar{D}^{0}D_s^{*+} + \bar{D}^{*0}D_s^+)]}<0.03$ at 90\% confidence level is determined at 4.68 GeV, where the $Z_{cs}(3985)$ is observed as the most significant. The corresponding ratio for the $Z_c(3900)$ case is around 0.1~\cite{BESIII:2013qmu}, suggesting that the $Z_{cs}(3985)$ and $Z_c(3900)$ may not be very similar. And the suppression of $Z_{cs}(3985)$ decaying to $KJ/\psi$ supports assigning $Z_{cs}(3985)$ and $Z_{cs}(4000)$ as two different states~\cite{Meng:2021rdg}.

\section{Conclusion}
The BESIII experiment has presented the most precise measurement for the cross sections of $e^+e^-\to K^+ K^- J/\psi$ at c.m.~energies between 4 and 5 GeV. The discovery of a new decay mode $Y(4230)\to K^+K^- J/\psi$, as well as two new vector charmoniumlike states, $Y(4500)$ and $Y(4710)$, has been reported. 
With current BESIII data, no significant $Z_{cs}$ signals have been detected in the $e^+e^-\to K^+ K^- J/\psi$ process. Then a constraint on the ratio $\frac{\mathcal{B}[Z_{cs}(3985)^{+}\rightarrow K^+ J/\psi]}{\mathcal{B}[Z_{cs}(3985)^{+}\rightarrow (\bar{D}^{0}D_s^{*+} + \bar{D}^{*0}D_s^+)]}<0.03$ has been set, which provides important input into the characteristics of $Z_{cs}(3985)$.

Over the past years, the BESIII Collaboration has made remarkable progress in discovering new hadrons, conducting precise measurements, and bring novel insights into the charmonium spectrum.
The BESIII experiment will continue its active, as an upgrade project for the BEPCII collider is scheduled in the summer of 2024. The upgrade aims to increase luminosity at higher energies, and extend the $e^+e^-$ c.m.~energy to 5.6 GeV. Stay tuned for more thrilling outcomes from the BESIII experiment.

\section*{Acknowledgments}
The speaker would like to thank the BESIII Collaboration and the organizer of the $58^\mathrm{th}$ Rencontres de Moriond for the opportunity to present these results.
This work is supported in part by China Postdoctoral Science Foundation under Grant No.~2023M74210, and Project ZR2022JQ02 supported by Shandong Provincial Natural Science Foundation.


\section*{References}


\end{document}